\documentclass[conference]{IEEEtran}
\IEEEoverridecommandlockouts
\def\BibTeX{{\rm B\kern-.05em{\sc i\kern-.025em b}\kern-.08em
    T\kern-.1667em\lower.7ex\hbox{E}\kern-.125emX}}

\usepackage{amsmath,amssymb,amsfonts}
\usepackage{algorithm}
\usepackage{algorithmic}
\usepackage{graphicx}
\usepackage{textcomp}
\usepackage{url}
\usepackage{xcolor}
\usepackage{caption}
\usepackage{subcaption}
\usepackage{threeparttable, booktabs}
\usepackage{tabularx}
\usepackage{array}
\usepackage{booktabs}
\usepackage{hyperref}
\hypersetup{
    colorlinks=true,
    urlcolor=blue,
    linkcolor=blue,
    citecolor=blue
}
\definecolor{red}{rgb}{1.0, 0.341, 0.2}
\definecolor{blue}{rgb}{0.290, 0.561, 0.882}

\begin{document}

\title{Piano: A Multi-Constraint Pin Assignment-Aware Floorplanner}
% \author{\IEEEauthorblockN{Anomymous Authors}}
\author{\IEEEauthorblockN{Zhexuan Xu$^1$, Kexin Zhou$^1$, Jie Wang$^1$, Zijie Geng$^1$, \\
Siyuan Xu$^2$, Shixiong Kai$^2$, Mingxuan Yuan$^2$ and Feng Wu$^1$}
  $^{1}$University of Science and Technology of China\\
  $^{2}$Noah’s Ark Lab, Huawei\\
}

\maketitle

\begin{abstract}
Floorplanning is a critical step in VLSI physical design, increasingly complicated by modern constraints such as fixed-outline requirements, whitespace removal, and the presence of pre-placed modules. In addition, the assignment of pins on module boundaries significantly impacts the performance of subsequent stages, including detailed placement and routing. However, traditional floorplanners often overlook pin assignment with modern constraints during the floorplanning stage.
In this work, we introduce Piano, a floorplanning framework that simultaneously optimizes module placement and pin assignment under multiple constraints. Specifically, we construct a graph based on the geometric relationships among modules and their netlist connections, then iteratively search for shortest paths to determine pin assignments. This graph-based method also enables accurate evaluation of feedthrough and unplaced pins, thereby guiding overall layout quality.
To further improve the design, we adopt a whitespace removal strategy and employ three local optimizers to enhance layout metrics under multi-constraint scenarios. Experimental results on widely used benchmark circuits demonstrate that Piano achieves an average 6.81\% reduction in HPWL, a 13.39\% decrease in feedthrough wirelength, a 16.36\% reduction in the number of feedthrough modules, and a 21.21\% drop in unplaced pins, while maintaining zero whitespace.
\end{abstract}


\section{Introduction}\label{sec:intro}
Floorplanning is the first step in modern VLSI physical design as it needs to determine the shape and location of large circuit modules on a chip canvas, while assigning the pins to each module's boundary for inter-module connections, thereby laying the foundation for subsequent detailed placement and routing stages.

However, modern floorplanners often neglect pin assignment during the floorplanning stage, as they treat floorplanning and pin assignment separately and sequentially: modules are placed first, and pin assignment is conducted based on the resulting layout. This separation can yield suboptimal outcomes. If pin positions are determined without accounting for their impact on inter-module wiring, or if floorplanning disregards the flexibility of pin locations, the resulting wiring may become unnecessarily long or congested~\cite{FloorplanningwithPinAssignment}.

In addition, to better align with real-world physical design, modern floorplanning needs to address a variety of design constraints including fixed-outline~\cite{fixed-outline}, whitespace removal~\cite{jigsawplanner} and pre-placed modules (PPMs)~\cite{TOFU}.

\begin{figure}[t]
    \centering
    \includegraphics[width=\linewidth]{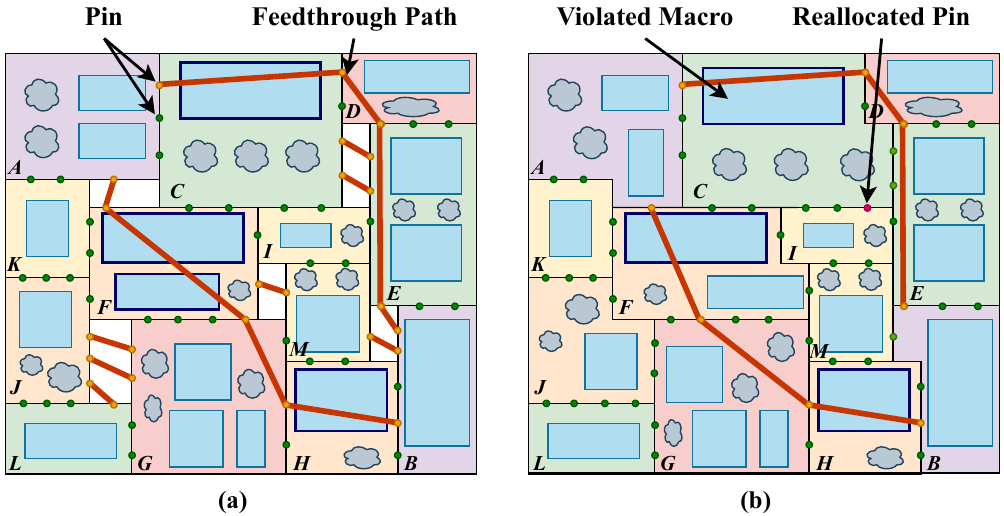}
    \caption{\textbf{Motivating example with pin assignment .} 
    (a) Layout with whitespace. (b) Layout with zero whitespace.
    }
    \label{fig:motivation}
\end{figure}

Figure~\ref{fig:motivation} provides a motivating example for this study.
Figure~\ref{fig:motivation}(a) demonstrates a tradition floorplanning solution with some whitespace.
All pins are assigned to module boundaries, with green pins representing nets connecting directly adjacent modules, thereby facilitating pin assignment on adjacent edges. Otherwise, feedthrough techniques are required for long-distance connections, as shown by the red path.

Figure~\ref{fig:motivation}(b) presents the layout after whitespace removal. Achieving a zero-whitespace layout offers multiple benefits. Notably, expanding module boundaries enables more effective pin reallocation, thereby reducing feedthrough and improving routability~\cite{TOFU}.
For example, suppose three nets connect \(M_C\) and \(M_I\). In Figure~\ref{fig:motivation}(a), the boundary of \(M_C\) cannot accommodate all three pins, resulting in one unplaceable pin. By contrast, reallocating whitespace to \(M_C\) increases its boundary length, allowing for the successful assignment of all pins, as shown in Figure~\ref{fig:motivation}(b).
Furthermore, this approach supports dynamic module adjustment, enhancing chip utilization and improving the placement of macros and standard cells within modules, which in turn mitigates the risk of macro violations.

Therefore, to address these challenges, we present \textbf{Piano}, a novel multi-constraint \underline{pi}n \underline{a}ssig\underline{n}ment-aware fl\underline{o}orplanner designed to co-optimize pin assignment during the floorplanning stage while adhering to multi modern constraints. To achieve this, we employ a three-stage framework. In the first stage, we use a wiremask to globally optimize the HPWL and a position mask to generate a legal floorplan. In the second stage, we construct a graph based on the netlist and the current geometric relationships among the modules on canvas. This enables us to assign pins accordingly and to compute the feedthrough and unplaced pin metrics. In the third stage, we apply a whitespace removal method to achieve a zero-whitespace design, while also employing three local optimizers to refine the canvas based on the metrics obtained in the second stage.

The key contributions of this paper are outlined as follows:
\begin{enumerate}
    \item \textbf{Concurrent Optimization:} We present a floorplanning framework that simultaneously optimizes module placement and pin assignment under multiple design constraints.
    \item \textbf{Graph-Based Pin Assignment:} We develop a graph-based algorithm for pin assignment, which also computes feedthrough and unplaced pin metrics.
    \item \textbf{Local Operators:} We utilize a simulated annealing framework with three local operators and a whitespace removal method to incrementally refine the floorplan.
    \item \textbf{Experimental Validation:} Experiments on benchmark circuits demonstrate that Piano reduces HPWL, feedthrough, and unplaced pins, while effectively assigning pins and handling PPMs.
\end{enumerate}

The remainder of this paper is organized as follows: Section~\ref{sec:previous work} surveys related work in the floorplanning field; Section~\ref{sec:preliminary} outlines the fundamental concepts of floorplanning and pin assignment; Section~\ref{sec:propose method} introduces our proposed methodology; Section~\ref{sec:experiment} presents the experimental results; and Section~\ref{sec:conclusion} provides the conclusions of our work.

\section{Previous Work}\label{sec:previous work}

Floorplanning methods have evolved significantly, with existing techniques broadly classified into three categories: discrete meta-heuristic, analytical solvers, and machine learning methods.

Heuristic methods, such as those based on geometric representations like the B*-Tree \cite{heuristic-SA}, Sequence Pair (SP) \cite{NP-hard}, and Corner Block List (CBL) \cite{CBL}, employ algorithms like simulated annealing (SA) to explore the solution space, often yielding near-optimal results. However, these methods typically require numerous iterations, which reduces computational efficiency.

Analytical methods, such as ePlace \cite{eplace} and UFO \cite{analytical-UFO}, define the objective function as a weighted average of wire length and density, following a two-stage process of global floorplanning and local legalization. During the global floorplanning stage, an initial solution with some overlaps is generated, which is then resolved in the local legalization stage. While efficient, these methods are constrained by the need for a differentiable objective function.

In recent years, machine learning (ML) methods have emerged as a promising approach. These methods excel at capturing features from the chip canvas and enhancing performance through continuous interaction with the environment via reinforcement learning \cite{DQN}. Despite their potential, machine learning (ML) methods face challenges due to limited training data, stemming from the proprietary nature of VLSI design. They also suffer from suboptimal sample efficiency, which limits model generalization and necessitates extensive training and fine-tuning.

Early work such as~\cite{FloorplanningwithPinAssignment} introduced a hierarchical algorithm to concurrently determine module placement and pin assignment. However, this method did not satisfy the modern constraints mentioned above. Over time, modern floorplanners have taken more constraints into account. ChipBench~\cite{benchmark} noted that although HPWL is computationally efficient, it frequently fails to align closely with the chip's final end-to-end performance. This observation has motivated current floorplanners to consider additional metrics, such as feedthrough and unplaced pins. FTAFP~\cite{FTAFP} has investigated feedthrough; however, its focus has generally been limited to feedthrough alone, without addressing the whitespace on canvas. Furthermore, TOFU~\cite{TOFU} proposed a method for near zero-whitespace while taking PPMs into consideration; however, it is used as a post-processing method and heavily relies on a high-quality initial solution. To address this issue, JigsawPlanner~\cite{jigsawplanner} proposed a strategy for achieving zero-whitespace floorplans; however, it often results in irregular module boundaries, posing challenges for placing macros within the modules. 
Moreover, these modern methods separate floorplanning and pin assignment, focusing solely on module placement.

\section{Preliminaries}\label{sec:preliminary}
\subsection{Floorplanning}

Floorplanning is the initial stage of the physical design process and serves as the critical link between the gate-level netlist and the actual physical layout on the chip.
Given a netlist \(\mathcal{N} = \{N_i \mid 1 \leq i \leq m\}\) and a set of modules \(\mathcal{M} = \{M_i \mid 1 \leq i \leq n\}\), each net \(N_i\) connects the required modules and the chip's I/O ports to enable inter-module communication. Each module \(M_i\) has a predefined area \(a_i\) and, if it is a soft block, an aspect ratio range. 
The objective of floorplanning is to optimally position and shape each module within the fixed chip canvas \((W, H)\)~\cite{TOFU}, aiming to minimize the intermediate metric, half-perimeter wirelength (HPWL), subject to modern constraints imposed by the design configuration.
The HPWL is defined as the sum of the half-perimeters of the bounding boxes for all nets in the netlist, as expressed by the following formula:
\begin{equation}
\text{HPWL} = \sum_{N_i \in \mathcal{N}} (w_i + h_i)
\label{eq:hpwl}
\end{equation}
where \(w_i\) and \(h_i\) represent the width and height of the bounding box for each net \(N_i\), respectively.

As mentioned above, floorplanning seeks to achieve a zero-whitespace solution while maintaining module regularity and minimizing module area variation. In this paper, module regularity refers to the total number of edge segments a module possesses and the module area variation rate (\textbf{AVR}) quantifies the difference between the predefined area and the final module area, calculated as:
\begin{equation}
\text{AVR}(i) = \frac{a_i - \text{CArea}(M_i)}{a_i}
\label{eq:AVR}
\end{equation}
where \(\text{CArea}(M_i)\) denotes the final canvas area of module \(M_i\). Noticing that the reason we incorporate AVR is to ensure the CArea of each module does not become significantly smaller than $a_i$. A reduced module area may cause illegal macro placement within it. In contrast, a slightly increased module area has limited impact. Therefore, we constrain the AVR of each module to be less than a predefined positive threshold.

\begin{figure*}[t]
    \centering
    \includegraphics[width=\textwidth]{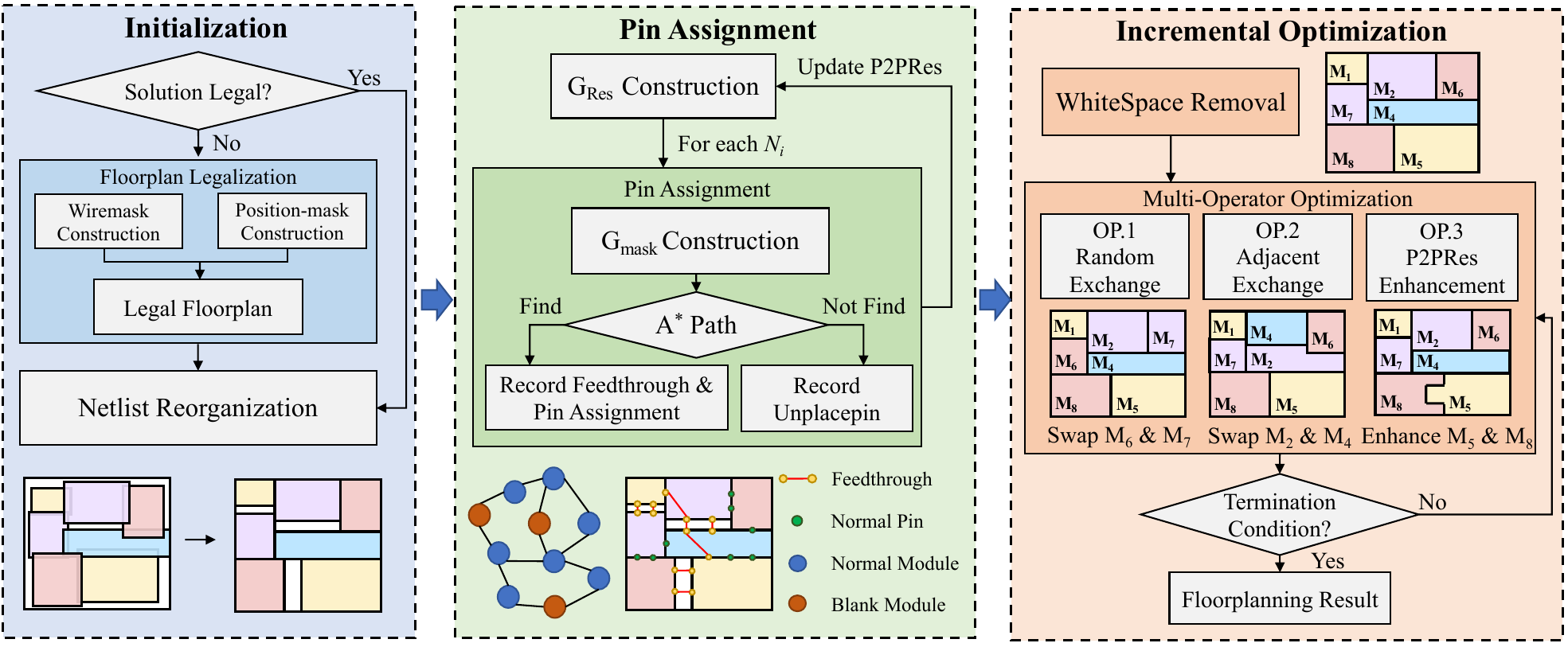}
    \caption{\textbf{Framework of Piano.} 
    Piano consists of three stages. The first stage performs legalization with global optimization. The second stage conducts pin assignment and feedthrough path detection. The third stage applies a whitespace removal method and three operators to perform local optimization.}
    \label{fig:overall flow}
\end{figure*}

\subsection{Pin Assignment}
In the context of physical design, the netlist describes how modules are interconnected. In practical floorplanning, the terminals of each net---represented as pins---are assigned along the boundaries of their respective modules, with a minimum spacing between pins, defined as \textbf{PinSpace}. We define the pin-to-pin connection resource (\textbf{P2PRes}) for a module pair as the maximum number of pins that can be assigned to their adjacent edge on the current canvas, given by:
\begin{equation}
\label{eq:p2pres}
\text{P2PRes}(M_i, M_j) = \left\lfloor \frac{\text{AvailEdge}_{ij}}{u} \right\rfloor
\end{equation}
where \(\text{AvailEdge}_{ij}\) represents the available adjacent edge length between modules \(M_i\) and \(M_j\), and \(u\) denotes the PinSpace. As pins are assigned, the available edge length decreases, reducing \(\text{P2PRes}\) accordingly.

If a net connects modules that are adjacent in the physical layout, their pins can be directly assigned on the adjacent edge boundary, facilitating data transmission between them. However, if the modules are non-adjacent, or if they are adjacent but the P2PRes has been fully utilized by other pins, feedthrough becomes necessary. Feedthrough is a through-module connection technique that enables long-distance communication between modules. A feedthrough path may enter one module and exit from another side, effectively routing through that module, although this may disrupt component placement within it.

Consider a net \(N = \{M_1, M_n\}\) that connects modules \(M_1\) and \(M_n\) through a feedthrough path \(\mathcal{P} = \{M_1, M_2, \ldots, M_n\}\), where \(n > 2\). We then define the feedthrough length (\(ft_{len}\)) and the number of feedthrough modules (\(ft_{num}\)) as follows:
\begin{equation}
ft_{len}(N) = \sum_{k=1}^{n-1} \text{Distance}(C_{M_k}, C_{M_{k+1}})\quad ft_{num}(N) = n - 2
\label{eq:ftlen}
\end{equation}
where \(\text{Distance}(C_{M_i}, C_{M_j})\) represents the Euclidean distance between the centers of modules \(M_i\) and \(M_j\). Using Euclidean distance provides a more accurate estimate of the actual length of each feedthrough net compared to the bounding-box approximation used in HPWL. We further define \textbf{FTlen} and \textbf{FTnum} as the average values of \(ft_{len}\) and \(ft_{num}\) across all nets, respectively.

While feedthrough enables connections over longer distances, it requires additional feedthrough pins and may compromise the implementability of component placement within modules traversed by feedthrough paths. Therefore, it is preferable for modules connected by a net to be physically adjacent or proximate in order to minimize feedthrough.

If no suitable feedthrough path can be established on the current canvas, it indicates that the net cannot be properly routed under the current floorplan layout and the specified PinSpace constraint. In this case, the net is deemed illegal and is marked with an \textbf{Unplacepin}.

\section{Proposed Method}\label{sec:propose method}
\subsection{Overview}

Building on the previous discussion, we introduce Piano, a framework that performs pin assignment, as well as feedthrough and unplaced pin calculation. Additionally, Piano incorporates whitespace removal and three specialized local operators to optimize these metrics, enabling it to serve as an incremental optimizer for any existing floorplanning algorithm.
As illustrated in Figure~\ref{fig:overall flow}, Piano contains three stages: (1) Initialization, (2) Pin Assignment, and (3) Incremental Optimization. The details of our algorithm are elaborated in Sections \ref{sec:legalization}, \ref{sec:graph}, and \ref{sec:operator}, respectively.

\subsection{Initialization}\label{sec:legalization}

In the initial stage, we apply a two-step preprocessing approach to the given circuit, consisting of floorplan legalization and netlist reorganization.

\begin{figure}[t]
   \centering
   \begin{subfigure}[t]{0.48\linewidth}
      \centering
      \includegraphics[width=\linewidth, page=1]{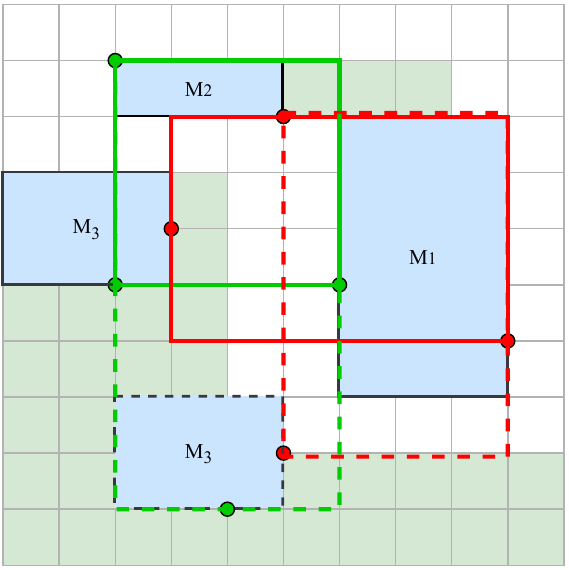}
      \caption{Position-mask}
      \label{fig:position-mask}
   \end{subfigure}
   \hfill
   \begin{subfigure}[t]{0.48\linewidth}
      \centering
      \includegraphics[width=\linewidth, page=2]{mask.pdf}
      \caption{Wiremask}
      \label{fig:wire-mask}
   \end{subfigure}
   \caption{\textbf{Visualization of Position-mask and Wiremask.} The position-mask identifies the legal grid, while the wiremask determines the HPWL increment when placing the module at each grid location.}
   \label{fig:mask}
\end{figure}

\subsubsection{\textbf{Floorplan Legalization}} 
First, we partition the chip canvas at the grid level, reformulating the original problem as a grid assignment task. In this task, each grid cell is assigned to a specific module with the goal of eliminating overlaps. Subsequently, we construct the \textit{Wiremask} and \textit{Position-mask} \cite{maskplace, wiremask-bbo} to globally optimize and legalize the initial solution. Note that our algorithm supports incremental optimization from a provided initial layout or from scratch. If the initial layout is legal and free of overlaps, this legalization step is skipped. However, if minor overlaps are exist in the initial solution, we employ the legalization step using these two masks to ensure a valid solution for the subsequent optimization stage. When starting from a raw netlist, we randomly generate multiple initial floorplans for all modules, followed by the legalization step, and choose the best floorplan with the least HPWL for global optimization.

The \textit{Wiremask} quantifies the increase in HPWL when placing the next module at each grid location, while the \textit{Position-mask} identifies legal grid positions that avoid overlaps with previously placed modules. Modules are processed sequentially in descending order \(a_i\). Each module \(M_i\) is relocated to the nearest available grid cell from its original position, minimizing the HPWL increase as determined by the element-wise product of the \textit{Wiremask} and \textit{Position-mask}, while ensuring that no overlaps occur.

As shown in Figure~\ref{fig:mask}, consider \(M_3\) as the current module to be placed. The \textit{Position-mask} is a matrix that highlights legal grid positions (marked in green) for the bottom-left grid cell of \(M_3\), while the \textit{Wiremask} is a matrix indicating the HPWL increase associated with placing \(M_3\) at each grid position. The red and green solid boxes respectively represent the bounding boxes of two nets.

\subsubsection{\textbf{Netlist Reorganization}} 
Second, to accurately compute feedthrough, we reorganize the netlist by decomposing nets with three or more connections (e.g., \(N_i = \{M_I, M_{O_1}, M_{O_2}, \ldots\}\)) into multiple two-module nets based on their I/O configurations, given that each net \(N_i\) has at most one input pin. For example, we create nets such as \(N_{i_1} = \{M_I, M_{O_1}\}, N_{i_2} = \{M_I, M_{O_2}\}, \ldots\), where \(M_I\) denotes the signal-generating module and \(M_{O_k}\) (\(k = 1, 2, \ldots\)) represents the signal-receiving modules. This restructuring facilitates parallel signal propagation from input to output pins, avoiding excessively long paths that could violate timing constraints if signals were propagated sequentially. 

Furthermore, each connected whitespace region is treated as a distinct blank module to ensure that each pin is assigned to the boundary along the adjacent edge between two modules.

\subsection{Pin Assignment}
\label{sec:graph}

In the second stage, given a legal layout, we construct a graph \(G_{\text{Res}}\) based on the P2PRes metric to facilitate pin assignment. This process also involves computing the feedthrough and unplaced pin metrics to support subsequent local optimization. The detailed steps are outlined in Algorithm~\ref{alg:graph_construction}.

\begin{figure*}[t]
\centering
\includegraphics[width=\textwidth]{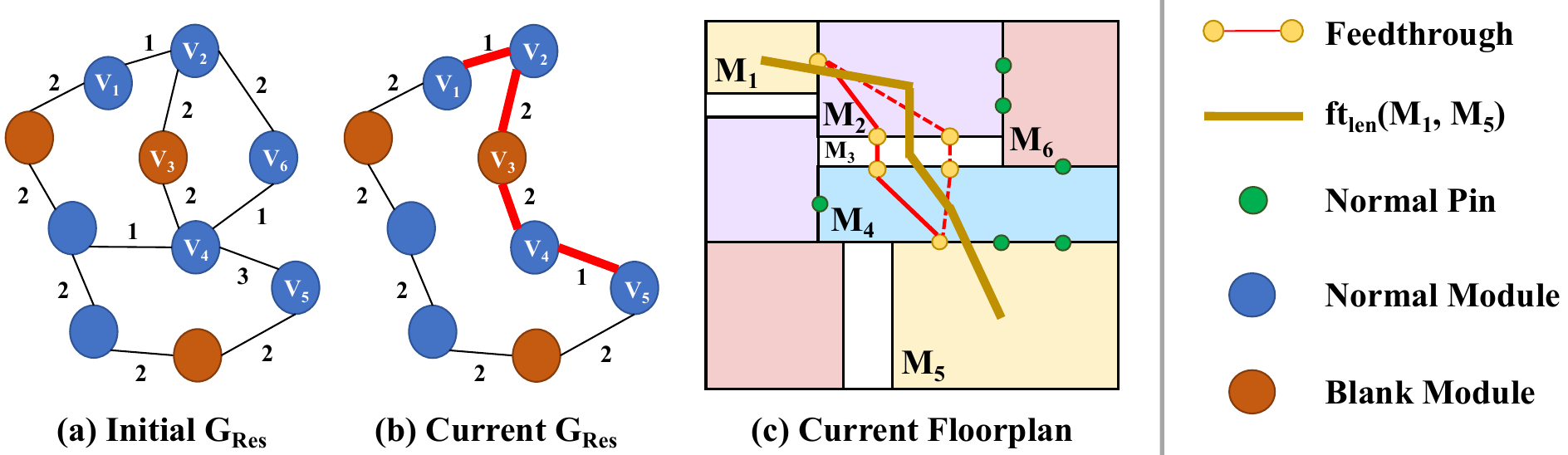}
\caption{\textbf{Visualization of Graph Construction and Pin Assignment.} (a) shows the initial \(G_{\text{Res}}\), while (b) and (c) illustrate \(G_{\text{Res}}\) and the corresponding floorplan during the routing of net \(N = \{M_{1}, M_{5}\}\).}
\label{fig:graph}
\end{figure*}

Each node \(\mathcal{V}_i\) in \(G_{\text{Res}}\) represents a module \(M_i\), with the edge weight \(\mathcal{E}_{ij}\) between nodes \(\mathcal{V}_i\) and \(\mathcal{V}_j\) corresponding to the current \(\text{P2PRes}(M_i, M_j)\), as defined in Equation~\ref{eq:p2pres}. An edge is removed if \(\text{P2PRes}(M_i, M_j)\) falls below the PinSpace threshold. 
Figure~\ref{fig:graph}(a) shows the initial \(G_{\text{Res}}\) corresponding to the floorplan in Figure~\ref{fig:graph}(c), before pin assignment.
For each adjacent edge between two modules, we partition it uniformly based on the PinSpace constraint, which corresponds to the actual physical locations of pins along the module boundaries.

We process each net in the following order. First, we traverse the nets connecting two adjacent modules, allowing pins to be assigned directly on their adjacent edge and eliminating the need for feedthrough across other modules. Next, we process the nets that could not be assigned in the first pass, indicating that although they connect adjacent modules, the P2PRes is insufficient and feedthrough must be introduced. Finally, we handle the nets connecting two non-adjacent modules.
In each iteration, we identify a physical path representing the actual routing of the net and assign pins accordingly.

To efficiently assign pins of that net (e.g., \(N_i = \{M_{i_1}, M_{i_n}\}\)), we construct an additional graph \(G_{\text{mask}}\) derived from \(G_{\text{Res}}\). The graph \(G_{\text{mask}}\) shares the same vertices as \(G_{\text{Res}}\), and an edge \(\mathcal{E}_{ij}'\) exists between vertices \(\mathcal{V}_i'\) and \(\mathcal{V}_j'\) in \(G_{\text{mask}}\) if and only if the corresponding edge \(\mathcal{E}_{ij}\) in \(G_{\text{Res}}\) has a positive weight, indicating available P2PRes. In this case, the edge weight \(\mathcal{E}_{ij}'\) is defined as the Euclidean distance between the geometric centers of modules \(M_i\) and \(M_j\). We then apply the \(A^*\) algorithm to find the shortest path between the vertex pair \((\mathcal{V}_{i_1}', \mathcal{V}_{i_n}')\) in \(G_{\text{mask}}\). The heuristic function of \(A^*\) is set to the Euclidean distance to the destination vertex.

\begin{algorithm}[t]
\caption{Graph Construction}
\label{alg:graph_construction}
\begin{algorithmic}[1]
\STATE \textbf{Input:} Legal Canvas $\mathcal{C}$, Reorganized Netlist $\mathcal{N}$
\STATE \textbf{Output:} $\mathcal{C}$ with pin assignment and related metrics
\STATE $\mathcal{V}_i \gets \text{Node}(M_i)$, $\mathcal{E}_{ij} \gets \text{P2PRes}(M_i, M_j)$
\STATE Construct graph $G_{\text{Res}} \gets \text{Graph}(\mathcal{V}, \mathcal{E})$ 
% \STATE Initialize $FT_{len} \gets 0$, $FT_{num} \gets 0$, $\text{Unplacepin} \gets 0$
\FOR{each net $N_i = \{M_{i_1}, M_{i_n}\} \in \mathcal{N}$}
    \STATE $\mathcal{V}' \gets \mathcal{V}$, $\mathcal{E}_{ij}' \gets \text{Distance}(M_i, M_j)*\text{Relu}(\mathcal{E}_{ij})$
    
    % \FOR{each edge $\mathcal{E}_{ij}$ in $G_{\text{Res}}$}
    %     \IF{$\mathcal{E}_{ij} > 0$}
    %         \STATE $\mathcal{E}'_{ij} \gets \text{EuclideanDistance}(M_i, M_j)$
    %     \ENDIF
    % \ENDFOR
    
    \STATE Construct graph $G_{\text{mask}} = \text{Graph}(\mathcal{V}', \mathcal{E}')$
    \STATE $\mathcal{P} \gets A^*(G_{\text{mask}}, \mathcal{V}_{i_1}', \mathcal{V}_{i_n}')$
    \IF{$\mathcal{P}$ exists} 
        \IF{$\mathcal{P} = \{\mathcal{V}_{i_1}', \mathcal{V}_{i_n}'\}$}
            \STATE \text{PinAssign}($\mathcal{P}$)
            \STATE $\mathcal{E}_{i_1,i_n} \gets \mathcal{E}_{i_1,i_n} - 1$
        \ELSE
            \STATE $\mathcal{P} = \{ \mathcal{V}_{i_1}', \mathcal{V}_{i_2}', \mathcal{V}_{i_3}', \dots, \mathcal{V}_{i_n}'\}$
            \STATE \text{BeamAssign}($\mathcal{P}$)
            \STATE $\mathcal{L} \gets 0$
            \FOR{$k$ from $1$ to $n-1$}
                \STATE $\mathcal{E}_{i_k,i_{k+1}} \gets \mathcal{E}_{i_k,i_{k+1}} - 1$
                \STATE $\mathcal{L} \gets \mathcal{L} + \mathcal{E}'_{i_k,i_{k+1}}$
            \ENDFOR
            \STATE FTlen $\gets \left( (i-1) \times \text{FTlen} + \mathcal{L} \right) \big/ i$
            \STATE $\text{FTnum} \gets \left( (i-1) \times \text{FTnum} + (n - 2) \right) \big/ i$
        \ENDIF
    \ELSE 
        \STATE $\text{Unplacepin} \gets \text{Unplacepin} + 1$
    \ENDIF
\ENDFOR
\STATE \textbf{Return:} $\mathcal{C}$, FTlen, FTnum, $\text{Unplacepin}$
\end{algorithmic}
\end{algorithm}

\subsubsection*{\textbf{Path Found}}
The following cases describe the pin assignment process when a path is identified:

\textbf{Case 1.1}: If the path involves only two modules---i.e., \(\mathcal{V}_{i_1}'\) and \(\mathcal{V}_{i_n}'\) are adjacent in \(G_{\text{mask}}\)---no feedthrough is required. Pins are directly assigned to the adjacent edge between \(M_{i_1}\) and \(M_{i_n}\), and the edge weight \(\mathcal{E}_{i_1,i_n}\) in \(G_{\text{Res}}\) is decremented by 1 to reflect the assigned pin.

\textbf{Case 1.2}: If the path \(\mathcal{P} = \{\mathcal{V}_{i_1}', \mathcal{V}_{i_2}', \mathcal{V}_{i_3}', \dots, \mathcal{V}_{i_n}'\}\) includes at least one additional module, pin assignment requires feedthrough, with the additional modules serving as feedthrough modules. Since each edge in \(\mathcal{P}\) exists in \(G_{\text{mask}}\), sufficient space remains on the adjacent edges between consecutive modules to allocate feedthrough pins. For each pair of adjacent modules \((\mathcal{V}_{i_k}', \mathcal{V}_{i_{k+1}}')\) along the path, we reduce the edge weight \(\mathcal{E}_{i_k,i_{k+1}}\) in \(G_{\text{Res}}\) by 1, indicating the need for additional feedthrough pin to establish the path.
Figure~\ref{fig:graph}(b) shows \(G_{\text{Res}}\) after the normal pins (marked in green) have been properly assigned. When routing the net \(N = \{M_{1}, M_{5}\}\), the corresponding \(A^*\) path is highlighted by the bold red line in Figure~\ref{fig:graph}(b), although \(G_{\text{mask}}\) is not explicitly shown.

% \begin{algorithm}[t]
% \caption{Pin Assignment}
% \label{alg:pin assignment}
% \begin{algorithmic}[1]
% \STATE \textbf{Input:} Module $M_i$, $M_j$ and the pin's type $type$
% \STATE \textbf{Ensure:} Set the specific type pin on their adjacent edge
% \STATE $\text{PinSet} \gets \{ \text{All pins corresponding to P2PRes($M_i$, $M_j$)}\}$
% \STATE Sort the PinSet by $\text{Distance}(M_i, pin) + \text{Distance}(pin, M_j)$
% \FOR{each pin $\in \text{PinSet}$}
%     \IF {pin is not available}
%         \STATE continue
%     \ELSE
%         \STATE pin's type $\gets type$ 
%         \STATE \textbf{return}
%     \ENDIF
% \ENDFOR
% \end{algorithmic}
% \end{algorithm}

After determining the modules involved in a feedthrough path, the next step is to identify the precise pin locations. When assigning pins for the net \(N = \{M_{1}, M_{5}\}\), there are two available pin locations on each of the module pairs \((M_2, M_3)\) and \((M_3, M_4)\), resulting in four potential physical feedthrough paths. Two of these paths are illustrated in Figure~\ref{fig:graph}(c) using solid and dashed red lines.

Our objective is to find the physical feedthrough path with the shortest Euclidean length. However, the complexity of this problem grows exponentially with the number of modules in the path, as long as multiple available pin positions exist for each module pair. Noting that the set of modules on the path \(\mathcal{P}\) is fixed, we adopt a beam search strategy to solve this problem, balancing between computational efficiency and solution quality. 
We begin at the signal-generating module corresponding to the net represented by \(\mathcal{P}\) and maintain the top-\(k\) shortest paths identified so far. For each subsequent module, we evaluate all available pins, extend each of the current beam paths with these candidate pins, and update the stored top-\(k\) paths accordingly. This process is repeated until a complete path is constructed.

However, the \(ft_{len}\) metric for each feedthrough net is defined as the sum of the Euclidean distances between the geometric centers of adjacent modules along the feedthrough path \(\mathcal{P}\), as specified in Equation~\ref{eq:ftlen}, rather than the actual physical distance computed above. This is because the actual feedthrough routing depends on the internal component placement within modules, which is unavailable at the current floorplanning stage. Therefore, using module centers provides a valid approximation in expectation. For the net \(N = \{M_{1}, M_{5}\}\), \(ft_{len}(N)\) corresponds to the length of the brown line segments shown in Figure~\ref{fig:graph}(c).

\subsubsection*{\textbf{Path Not Found}}
If no path exists---i.e., \( \mathcal{V}_{i_1}' \) and \( \mathcal{V}_{i_n}' \) are not connected in \( G_{\text{mask}} \)---the net is considered illegal in the current layout, as no feedthrough can connect the modules. Consequently, the number of Unplacepin is incremented, and feedthrough-related metrics remain unchanged.

\subsection{Incremental Optimization}
\label{sec:operator}

In the third stage, Piano employs a two-step incremental optimization process. First, whitespace removal is applied to the layout. Then, three specialized operators are introduced to locally optimize modules along the feedthrough paths, thereby enhancing overall layout quality.

\subsubsection{\textbf{Whitespace Removal}}
\label{sec:WS-remove}

In the first step, we optimally allocate the remaining whitespace. To enhance P2PRes between modules, we first perform in-place rectangular expansion on each module, prioritizing those with a higher number of unplaced pins. This process ensures that each module retains its rectangular shape after expansion. However, the rectangular constraint may leave some whitespace unallocated. For each contiguous whitespace region, we assign it to the adjacent module that minimizes FTlen.

\subsubsection{\textbf{Multi-Operator Optimization}}

Following whitespace removal, we perform local optimization based on the feedthrough paths identified in Stage~2. To achieve this, we introduce three operators, each governed by specific constraints.

\textit{Operator 1: Random Exchange.} For any two modules \(M_i\) and \(M_j\), not necessarily adjacent, this operator permits their exchange on the canvas provided that the area variation rate (defined in Equation~\ref{eq:AVR}) after the swap remains within a permissible threshold.

\textit{Operator 2: Adjacent Exchange.} For two adjacent modules \(M_i\) and \(M_j\), which may differ in canvas area (\(\text{CArea}\)), this operator facilitates their exchange by dynamically adjusting their adjacent boundary to preserve their original areas. To achieve this, we maintain a boundary queue to track the grids along the adjacent edge. In each iteration after the swap, boundary grids are reassigned to the smaller module until its initial area is restored. During this process, we ensure that the boundary grids are either entirely altered or remain unchanged, thereby preserving the original boundary pattern of the modules.

\textit{Operator 3: P2PRes Enhancement.} For two adjacent modules \(M_i\) and \(M_j\), this operator aims to increase their P2PRes while maintaining module regularity. To this end, slots are introduced along their adjacent edge to extend its length, thereby improving their P2PRes. Given a straight segment \(AB\), we identify its trisection points \(C_1\) and \(C_2\), and extend the sub-segment \(C_1C_2\) in one random direction by a length equal to \(|C_1C_2|\), while preserving connectivity between modules. To compensate for the resulting area change, the slot is shifted in the opposite direction of its extension to restore the original area. However, each slot introduces jagged edges, increasing boundary irregularity and potentially harming component placement within the module. To balance this, we impose a hyperparameter that limits the total number of edge segments per module.

The primary goals at this stage are: (1) to position modules connected by the same net adjacently, thereby minimizing feedthrough; and (2) to increase module boundaries in order to reduce the number of unplaced pins.
To this end, we traverse each feedthrough path after whitespace removal (\(\mathcal{P} = \{\mathcal{V}_{i_1}', \mathcal{V}_{i_2}', \mathcal{V}_{i_3}', \dots, \mathcal{V}_{i_n}'\}\)) and iteratively apply Operators 1 and 2 to reduce the distance between \(M_{i_1}\) and \(M_{i_n}\). If \(M_{i_1}\) and \(M_{i_n}\) become adjacent but their shared adjacent edge lacks sufficient pin capacity, Operator 3 is utilized to enhance their P2PRes. Under this framework, we adopt a simulated annealing strategy to apply the operators, facilitating broader exploration of the solution space, effectively avoiding local minima, and yielding a more comprehensive and robust optimization outcome.

\newcolumntype{C}{>{\centering\arraybackslash}X}
\begin{table*}[t]
\centering
\caption{The main results for the HPWL, FTlen, FTnum, Unplacepin, and runtime (RT, in seconds) metrics on the GSRC and MCNC benchmarks. The \textbf{\textcolor{red}{best}} results are highlighted in bold red, and the \textcolor{blue}{\underline{second-best}} results are underlined in blue.}
\label{tab:main}
\begin{tabularx}{\textwidth}{CCCCCCCCCC} % 10 columns, using custom C column type
\toprule
\textbf{Method} &
  \textbf{Metric} &
  \textbf{n10} &
  \textbf{n30} &
  \textbf{n50} &
  \textbf{n100} &
  \textbf{n200} &
  \textbf{n300} &
  \textbf{ami33} &
  \textbf{ami49} \\ 
\midrule
Corblivar &
  \begin{tabular}[c]{@{}c@{}}HPWL \\ FTlen \\ FTnum \\ Unplacepin \\ RT(s)\end{tabular} &
  \begin{tabular}[c]{@{}c@{}}47316\\ 101.64\\ 2.25\\ 27\\ \textbf{\textcolor{red}{0.04}}\end{tabular} &  % n10
  \begin{tabular}[c]{@{}c@{}}133196\\ 72.57\\ 3.12\\ 88\\ \textbf{\textcolor{red}{0.34}}\end{tabular} & % n30
  \begin{tabular}[c]{@{}c@{}}160333\\ 69.75\\ 3.88\\ 250\\ \textbf{\textcolor{red}{0.98}}\end{tabular} & % n50
  \begin{tabular}[c]{@{}c@{}}260674\\ 49.69\\ 3.92\\ 494\\ \textcolor{blue}{\underline{4.83}}\end{tabular} & % n100
  \begin{tabular}[c]{@{}c@{}}520883\\ 43.46\\ 4.86\\ 1232\\ 26.97 \end{tabular} & % n200
  \begin{tabular}[c]{@{}c@{}}714560\\ 38.15\\ 5.32\\ 1671\\ 69.53\end{tabular} & % n300
  \begin{tabular}[c]{@{}c@{}}85441\\ 69.00\\ 2.55\\ 244\\ \textbf{\textcolor{red}{0.37}}\end{tabular} & % ami33
  \begin{tabular}[c]{@{}c@{}}1666454\\ 91.80\\ 5.08\\ 442\\ \textbf{\textcolor{red}{0.94}}\end{tabular} \\ % ami49
\midrule
TOFU &
  \begin{tabular}[c]{@{}c@{}}HPWL \\ FTlen \\ FTnum \\ Unplacepin \\ RT(s)\end{tabular} &
  \begin{tabular}[c]{@{}c@{}}\textcolor{blue}{\underline{35213}}\\ 96.22\\ 2.13\\ 28\\ \textcolor{blue}{\underline{1.35}}\end{tabular} & % n10
  \begin{tabular}[c]{@{}c@{}}\textcolor{red}{\textbf{109189}}\\ 76.87\\ 3.56\\ 104\\ \textcolor{blue}{\underline{0.97}}\end{tabular} & % n30
  \begin{tabular}[c]{@{}c@{}}\textcolor{blue}{\underline{147650}}\\ \textcolor{blue}{\underline{64.95}}\\ \textcolor{blue}{\underline{3.51}}\\ 258\\ \textcolor{blue}{\underline{1.55}}\end{tabular} & % n50
  \begin{tabular}[c]{@{}c@{}}250942\\ \textcolor{blue}{\underline{45.99}}\\ \textcolor{blue}{\underline{3.65}}\\ 510\\ \textbf{\textcolor{red}{2.58}}\end{tabular} & % n100
  \begin{tabular}[c]{@{}c@{}}484507\\ 48.83\\ 5.65\\ 1288\\ \textbf{\textcolor{red}{3.81}}\end{tabular} & % n200
  \begin{tabular}[c]{@{}c@{}}712547\\ 46.72\\ 6.76\\ 1725\\ \textbf{\textcolor{red}{8.21}}\end{tabular} & % n300
  \begin{tabular}[c]{@{}c@{}}\textcolor{blue}{\underline{71069}}\\ \textcolor{blue}{\underline{60.33}}\\ \textcolor{blue}{\underline{2.13}}\\ \textcolor{blue}{\underline{238}}\\ \textcolor{blue}{\underline{1.15}}\end{tabular} & % ami33
  \begin{tabular}[c]{@{}c@{}}1144620\\ 69.68\\ 3.71\\ 440\\ \textcolor{blue}{\underline{1.58}}\end{tabular} \\ % ami49
\midrule
Wiremask-EA &
  \begin{tabular}[c]{@{}c@{}}HPWL \\ FTlen \\ FTnum \\ Unplacepin \\ RT(s)\end{tabular} &
  \begin{tabular}[c]{@{}c@{}}38608\\ \textcolor{blue}{\underline{87.91}}\\ \textcolor{blue}{\underline{1.86}}\\ \textcolor{blue}{\underline{18}}\\ 24.96\end{tabular} & % n10
  \begin{tabular}[c]{@{}c@{}}120406\\ \textcolor{blue}{\underline{62.67}}\\ \textcolor{blue}{\underline{2.27}}\\ \textcolor{blue}{\underline{72}}\\ 45.94\end{tabular} & % n30
  \begin{tabular}[c]{@{}c@{}}157501\\ 68.23\\ 3.59\\ \textcolor{blue}{\underline{230}}\\ 67.15\end{tabular} & % n50
  \begin{tabular}[c]{@{}c@{}}\textcolor{blue}{\underline{245036}}\\ 51.51\\ 4.03\\ \textcolor{blue}{\underline{487}}\\ 102.27\end{tabular} & % n100
  \begin{tabular}[c]{@{}c@{}}\textcolor{blue}{\underline{449048}}\\ \textcolor{blue}{\underline{33.81}}\\ \textcolor{blue}{\underline{3.54}}\\ \textcolor{blue}{\underline{1134}}\\ 168.84\end{tabular} & % n200
  \begin{tabular}[c]{@{}c@{}}\textcolor{blue}{\underline{621293}}\\ \textcolor{blue}{\underline{34.07}}\\ \textcolor{blue}{\underline{4.70}}\\ \textcolor{blue}{\underline{1559}}\\ 234.60\end{tabular} & % n300
  \begin{tabular}[c]{@{}c@{}}77137\\ 78.28\\ 3.83\\ 240\\ 49.78\end{tabular} & % ami33
  \begin{tabular}[c]{@{}c@{}}\textcolor{blue}{\underline{1098393}}\\ \textcolor{blue}{\underline{59.69}}\\ \textcolor{blue}{\underline{2.72}}\\ \textcolor{blue}{\underline{384}}\\ 51.31\end{tabular} \\ % ami49
\midrule
Piano &
  \begin{tabular}[c]{@{}c@{}}HPWL \\ FTlen \\ FTnum \\ Unplacepin \\ RT(s)\end{tabular} &
  \begin{tabular}[c]{@{}c@{}}
    \textbf{\textcolor{red}{34435}}\\ 
    \textbf{\textcolor{red}{66.23}}\\ 
    \textbf{\textcolor{red}{1.26}}\\ 
    \textbf{\textcolor{red}{1}}\\ 
    1.51
  \end{tabular} & % n10
  \begin{tabular}[c]{@{}c@{}}
    \textcolor{blue}{\underline{109938}}\\ 
    \textbf{\textcolor{red}{58.01}}\\ 
    \textbf{\textcolor{red}{2.21}}\\ 
    \textbf{\textcolor{red}{45}}\\ 
    2.84
  \end{tabular} & % n30
  \begin{tabular}[c]{@{}c@{}}
    \textbf{\textcolor{red}{138968}}\\ 
    \textbf{\textcolor{red}{50.99}}\\ 
    \textbf{\textcolor{red}{2.43}}\\ 
    \textbf{\textcolor{red}{210}}\\ 
    4.45
  \end{tabular} & % n50
  \begin{tabular}[c]{@{}c@{}}
    \textbf{\textcolor{red}{226381}}\\ 
    \textbf{\textcolor{red}{41.12}}\\ 
    \textbf{\textcolor{red}{3.05}}\\ 
    \textbf{\textcolor{red}{436}}\\ 
    8.46
  \end{tabular} & % n100
  \begin{tabular}[c]{@{}c@{}}
    \textbf{\textcolor{red}{428513}}\\ 
    \textbf{\textcolor{red}{32.10}}\\ 
    \textbf{\textcolor{red}{3.34}}\\ 
    \textbf{\textcolor{red}{1110}}\\ 
    \textcolor{blue}{\underline{21.66}}
  \end{tabular} & % n200
  \begin{tabular}[c]{@{}c@{}}
    \textbf{\textcolor{red}{584168}}\\ 
    \textbf{\textcolor{red}{27.89}}\\ 
    \textbf{\textcolor{red}{3.63}}\\ 
    \textbf{\textcolor{red}{1526}}\\ 
    \textcolor{blue}{\underline{41.93}}
  \end{tabular} & % n300
  \begin{tabular}[c]{@{}c@{}}
    \textbf{\textcolor{red}{61460}}\\ 
    \textbf{\textcolor{red}{50.88}}\\ 
    \textbf{\textcolor{red}{1.90}}\\ 
    \textbf{\textcolor{red}{228}}\\ 
    2.81
  \end{tabular} & % ami33
  \begin{tabular}[c]{@{}c@{}}
    \textbf{\textcolor{red}{934285}}\\ 
    \textbf{\textcolor{red}{57.27}}\\ 
    \textbf{\textcolor{red}{2.47}}\\ 
    \textbf{\textcolor{red}{345}}\\ 
    4.50
  \end{tabular} \\ % ami49
\bottomrule
\end{tabularx}
\end{table*}

\newcolumntype{C}{>{\centering\arraybackslash}X}
\begin{table*}[t]
\centering
\caption{The main results for the incremental optimization of Piano, applied to Corblivar's initial solution. The \textbf{\textcolor{red}{best}} results are highlighted in bold red.
}
\label{tab:incremental optimize}
\begin{tabularx}{\textwidth}{CCCCCCCCCC} % 10 columns, using custom C column type
\toprule
\textbf{Method} &
  \textbf{Metric} &
  \textbf{n10} &
  \textbf{n30} &
  \textbf{n50} &
  \textbf{n100} &
  \textbf{n200} &
  \textbf{n300} &
  \textbf{ami33} &
  \textbf{ami49} \\ 
\midrule
Corblivar &
  \begin{tabular}[c]{@{}c@{}}HPWL \\ FTlen \\ FTnum \\ Unplacepin \\ RT(s)\end{tabular} &
  \begin{tabular}[c]{@{}c@{}}47316\\ 101.64\\ 2.25\\ 27\\ \textbf{\textcolor{red}{0.04}}\end{tabular} &  % n10
  \begin{tabular}[c]{@{}c@{}}133196\\ 72.57\\ 3.12\\ 88\\ \textbf{\textcolor{red}{0.34}}\end{tabular} & % n30
  \begin{tabular}[c]{@{}c@{}}160333\\ 69.75\\ 3.88\\ 250\\ \textbf{\textcolor{red}{0.98}}\end{tabular} & % n50
  \begin{tabular}[c]{@{}c@{}}260674\\ 49.69\\ 3.92\\ 494\\ \textbf{\textcolor{red}{4.83}}\end{tabular} & % n100
  \begin{tabular}[c]{@{}c@{}}520883\\ 43.46\\ 4.86\\ 1232\\ 26.97 \end{tabular} & % n200
  \begin{tabular}[c]{@{}c@{}}714560\\ 38.15\\ 5.32\\ 1671\\ 69.53\end{tabular} & % n300
  \begin{tabular}[c]{@{}c@{}}85441\\ 69.00\\ 2.55\\ 244\\ \textbf{\textcolor{red}{0.37}}\end{tabular} & % ami33
  \begin{tabular}[c]{@{}c@{}}1666454\\ 91.80\\ 5.08\\ 442\\ \textbf{\textcolor{red}{0.94}}\end{tabular} \\ % ami49
\midrule
Piano &
  \begin{tabular}[c]{@{}c@{}}HPWL \\ FTlen \\ FTnum \\ Unplacepin \\ RT(s)\end{tabular} &
  \begin{tabular}[c]{@{}c@{}}
    \textbf{\textcolor{red}{41324}}\\ 
    \textbf{\textcolor{red}{69.69}}\\ 
    \textbf{\textcolor{red}{3}}\\ 
    \textbf{\textcolor{red}{1.35}}\\ 
    1.05
  \end{tabular} & % n10
  \begin{tabular}[c]{@{}c@{}}
    \textbf{\textcolor{red}{113366}}\\ 
    \textbf{\textcolor{red}{54.06}}\\ 
    \textbf{\textcolor{red}{2.00}}\\ 
    \textbf{\textcolor{red}{59}}\\ 
    1.89
  \end{tabular} & % n30
  \begin{tabular}[c]{@{}c@{}}
    \textbf{\textcolor{red}{147602}}\\ 
    \textbf{\textcolor{red}{55.71}}\\ 
    \textbf{\textcolor{red}{2.62}}\\ 
    \textbf{\textcolor{red}{210}}\\ 
    3.63
  \end{tabular} & % n50
  \begin{tabular}[c]{@{}c@{}}
    \textbf{\textcolor{red}{230526}}\\ 
    \textbf{\textcolor{red}{39.60}}\\ 
    \textbf{\textcolor{red}{2.88}}\\ 
    \textbf{\textcolor{red}{452}}\\ 
    6.24
  \end{tabular} & % n100
  \begin{tabular}[c]{@{}c@{}}
    \textbf{\textcolor{red}{432028}}\\ 
    \textbf{\textcolor{red}{33.26}}\\ 
    \textbf{\textcolor{red}{3.53}}\\ 
    \textbf{\textcolor{red}{1106}}\\ 
    \textbf{\textcolor{red}{17.92}}
  \end{tabular} & % n200
  \begin{tabular}[c]{@{}c@{}}
    \textbf{\textcolor{red}{624331}}\\ 
    \textbf{\textcolor{red}{31.14}}\\ 
    \textbf{\textcolor{red}{4.16}}\\ 
    \textbf{\textcolor{red}{1549}}\\ 
    \textbf{\textcolor{red}{34.15}}
  \end{tabular} & % n300
  \begin{tabular}[c]{@{}c@{}}
    \textbf{\textcolor{red}{72220}}\\ 
    \textbf{\textcolor{red}{80.95}}\\ 
    \textbf{\textcolor{red}{3.75}}\\ 
    \textbf{\textcolor{red}{241}}\\ 
    2.40
  \end{tabular} & % ami33
  \begin{tabular}[c]{@{}c@{}}
    \textbf{\textcolor{red}{996544}}\\ 
    \textbf{\textcolor{red}{60.28}}\\ 
    \textbf{\textcolor{red}{2.44}}\\ 
    \textbf{\textcolor{red}{363}}\\ 
    3.96
  \end{tabular} \\ % ami49
\bottomrule
\end{tabularx}
\end{table*}

\section{Experimental Results}\label{sec:experiment}

\subsection{Benchmarks and Settings}

In this section, we evaluate Piano against current state-of-the-art (SOTA) methods to demonstrate its effectiveness. All experiments were conducted on a single machine equipped with an Intel Xeon E5-2667 v4 CPU operating at 3.20\,GHz.

We evaluated Piano using the widely recognized MCNC~\cite{mcnc} and GSRC~\cite{gsrc} datasets, which include module counts ranging from 10 to 300, effectively covering both small-scale and large-scale designs. For comparative analysis, we selected three algorithms:  

\textbf{Corblivar~\cite{corb}}: An efficient simulated annealing-based floorplanning algorithm, renowned for its robust performance in the open-source community.

\textbf{TOFU~\cite{TOFU}}: An incremental optimization algorithm characterized by its exceptionally short runtime and effective whitespace removal. Since its implementation details are not publicly available, we obtained the layout results directly from the authors and integrated them into our framework to evaluate feedthrough and unplaced pin metrics.

\textbf{Wiremask-EA~\cite{wiremask-bbo}}: Originally developed for macro placement, but adapted for floorplanning due to its effectiveness in minimizing HPWL.

% Notably, Piano can function both as an incremental optimizer and as a floorplanner starting from scratch. To fully demonstrate the capabilities of our algorithm and mitigate the influence of initial solutions, we evaluated Piano starting from the original netlist. 

Notably, since only TOFU and Piano incorporate whitespace removal, we applied our whitespace removal technique to the final solutions of Corblivar and Wiremask-EA to ensure a fair comparison of pin assignment.

Performance was assessed using five evaluation metrics: HPWL, FTlen, FTnum, Unplacepin, and runtime (RT). Although HPWL is a standard metric in floorplanning, it does not account for pin assignment. To address this limitation, we proposed a pin assignment method and introduced FTlen and FTnum to analyze feedthrough characteristics, alongside Unplacepin to quantify the number of pins that could not be assigned. We did not report the whitespace area ratio, as our whitespace removal technique ensures zero whitespace in the final layouts of all methods except TOFU, which achieves a whitespace ratio of less than 1\%. In addition, we excluded the area variation rate (AVR) and module regularity, as these are treated as hyperparameters in our approach, adjustable based on design-specific configuration requirements. 

\textbf{Model Settings:} We set the grid size to \(224 \times 224\) for Piano and Wiremask-EA. PinSpace is defined as the ratio of the average perimeter of each module to the average number of nets connected per module. The AVR was set to 5\%, and the maximum edge count of each module was limited to 20. The value of \(k\) in the beam search algorithm is set to 5. The SA strategy in incremental optimization stage was configured with an initial temperature \(T_{\text{init}} = 100\), a final temperature \(T_{\text{end}} = 0.01\), and a cooling rate of 0.9. 
The objective function used in SA is a weighted combination of HPWL, FTlen, FTnum, and Unplacepin, with their weight ratio set to 1:50:2000:100, respectively.
Each approach was evaluated over five independent runs, and the mean values are reported.

\subsection{Main Results}

\begin{figure*}[t]
    \centering
    \begin{subfigure}{\textwidth}
        \centering
        % Subfigure 1: Page 1
        \begin{subfigure}[t]{0.24\textwidth}
            \centering
            \includegraphics[width=\linewidth, page=1]{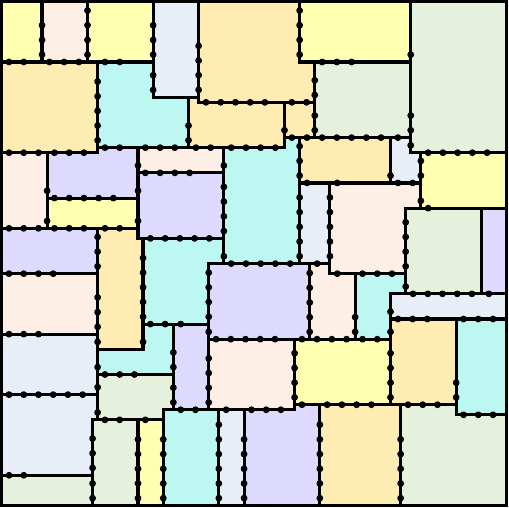}
            \caption{Corblivar}
            \label{fig:total_show_page1}
        \end{subfigure}
        \hfill
        % Subfigure 2: Page 2
        \begin{subfigure}[t]{0.24\textwidth}
            \centering
            \includegraphics[width=\linewidth, page=2]{main_result.pdf}
            \caption{Piano}
            \label{fig:total_show_page2}
        \end{subfigure}
        \hfill
        % Subfigure 3: Page 3
        \begin{subfigure}[t]{0.24\textwidth}
            \centering
            \includegraphics[width=\linewidth, page=3]{main_result.pdf}
            \caption{Piano with all Feedthrough}
            \label{fig:total_show_page3}
        \end{subfigure}
        \hfill
        % Subfigure 4: Page 4
        \begin{subfigure}[t]{0.24\textwidth}
            \centering
            \includegraphics[width=\linewidth, page=4]{main_result.pdf}
            \caption{Detail of \(M_0\)'s Feedthrough}
            \label{fig:total_show_page4}
        \end{subfigure}
    \end{subfigure}
    \caption{\textbf{Floorplan of GSRC \textit{n50} design.} \ref{fig:total_show_page1} displays the results of Corblivar after processing with our whitespace removal algorithm. \ref{fig:total_show_page3} and \ref{fig:total_show_page4} illustrate the feedthrough paths in the layout results by Piano.}
    \label{fig:total_show}
\end{figure*}

Table~\ref{tab:main} illustrates the performance of the discussed methods. In this test, Piano was configured to start from scratch, meaning it began the floorplanning from the initial netlist. Our algorithm consistently approaches optimal performance across feedthrough and Unplacepin metrics in eight datasets, while achieving nearly the best results in terms of HPWL. The results indicate that the Piano achieves an average reduction of 6.84\% in HPWL, 13.39\% in FTlen, and 16.36\% in FTnum across the datasets, along with a 21.21\% improvement in Unplacepin. However, since Piano considers pin assignments and explicitly assigns each pin, subsequently calculating feedthrough and Unplacepin metrics, its runtime is slightly longer compared to the fastest method. 
% Despite this, the runtime remains acceptable.

Meanwhile, to fully demonstrate the capabilities of our algorithm, we evaluated its performance as an incremental optimizer using the layout generated by Corblivar as the initial solution. The results, presented in Table~\ref{tab:incremental optimize}, indicate that our algorithm significantly enhances the solutions produced by Corblivar across all metrics, further highlighting its effectiveness. 
A direct comparison of the incremental optimization effectiveness between TOFU and Piano is not feasible, as the initial Corblivar solution used by TOFU is unknown, making a fair and reasonable comparison impossible.

We further compare the changes in five evaluation metrics when Piano is initialized from Corblivar versus from scratch. First, the runtime slightly decreases, as starting from scratch requires generating random solutions and performing global optimization and legalization using the \textit{Wiremask} and \textit{Position-mask}. Second, the HPWL shows a slight increase, which we attribute to the superior optimization capability of the \textit{Wiremask} in reducing HPWL. In contrast, FTlen, FTnum, and Unplacepin exhibit minimal variation, further demonstrating the effectiveness of Stage~2 and Stage~3 of Piano.

Figure~\ref{fig:total_show} displays the incremental optimization results produced by Piano based on the layout generated by Corblivar for the \textit{n50} dataset. We applied whitespace removal to Corblivar's solutions for better comparison, as shown in Figure~\ref{fig:total_show_page1}. The optimized results produced by Piano are presented in Figure~\ref{fig:total_show_page2}, where assigned pins are represented as black dots along the module boundaries. Figure~\ref{fig:total_show_page3} illustrates all feedthrough paths in the Piano layout, with paths sharing the same $ft_{num}$ colored identically; deeper colors indicate higher $ft_{num}$ values. 
To more clearly illustrate the metrics evaluated by Piano, we specifically highlight module \( M_0 \) in Figure~\ref{fig:total_show_page4}, displaying only the feedthrough paths associated with \( M_0 \).
Black pins on the boundaries of \(M_0\) signify direct connections to adjacent modules, eliminating the need for feedthrough. In contrast, colored pins denote paths requiring feedthrough. Notably, a feedthrough path connecting modules \(M_1\) and \(M_2\) passes through \(M_0\), designating \(M_0\) as a feedthrough module for this path. All other paths involving \(M_0\) terminate at this module, indicating \(M_0\) as a terminal endpoint for these nets.

\begin{figure}[!t]
    \centering
    % Subfigure 1: First Image
    \begin{subfigure}{0.48\columnwidth}
        \centering
        \includegraphics[width=\linewidth,page=1]{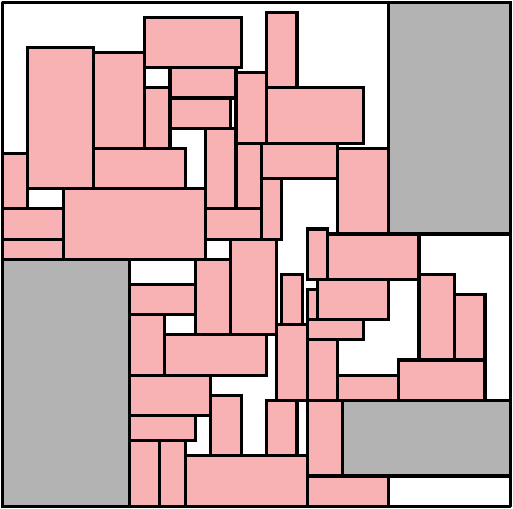}
        \caption{Corblivar}
        \label{fig:corblivar}
    \end{subfigure}
    \hfill
    % Subfigure 2: Second Image
    \begin{subfigure}{0.48\columnwidth}
        \centering
        \includegraphics[width=\linewidth,page=2]{preplace_result.pdf}
        \caption{Piano}
        \label{fig:stage1}
    \end{subfigure}
\caption{\textbf{PPMs Constraint Refinement.} (a) shows the initial solution generated by Corblivar, while (b) presents the incremental optimization under PPMs constraint performed by Piano. The PPMs are marked in grey.}
    \label{fig:PPM}
\end{figure}

\subsection{PPMs Constraint Refinement}
In this subsection, we demonstrate Piano's capability to handle pre-placed modules (PPMs), whose shapes and locations must remain fixed. Since the original netlists in the MCNC and GSRC datasets do not specify PPMs constraint, we randomly selected a subset of modules collectively occupying 30\% of the total area and designated them as PPMs, in order to better align with modern EDA design requirements. Discussions on PPMs must be based on a legal floorplanning solution. We initially used Corblivar, which does not support PPMs constraints, to generate a legal layout. Subsequently, we fixed the selected PPMs onto the canvas and applied Piano for further whitespace removal and pin assignment. Experimental results, shown in Figure~\ref{fig:PPM}, illustrate the effectiveness of our approach even under PPMs constraint.

\section{Conclusion}\label{sec:conclusion}
In this paper, we challenge existing HPWL-based floorplanning approaches by introducing Piano, a pin assignment-aware floorplanning framework capable of handling multiple constraints and serving as an incremental optimizer for any existing algorithm. We propose a pin assignment method incorporating feedthrough and unplaced pin models that better align with practical EDA requirements. Experimental results demonstrate that our algorithm outperforms SOTA methods in overall performance, offering a novel perspective for future research on pin assignment during the floorplanning stage.

\bibliographystyle{ieeetr}
\bibliography{ref}

\end{document}